\documentclass[fleqn,usenatbib]{mnras}

\usepackage{newtxtext,newtxmath}

\usepackage[T1]{fontenc}

\DeclareRobustCommand{\VAN}[3]{#2}
\let\VANthebibliography\thebibliography
\def\thebibliography{\DeclareRobustCommand{\VAN}[3]{##3}\VANthebibliography}

\usepackage{graphicx}	
\usepackage{amsmath}	
\usepackage{multirow}
\usepackage{booktabs}
\usepackage{hyperref}
\usepackage{ulem}
\newcommand{\orcid}[1]{\href{https://orcid.org/#1}{\includegraphics[width=8pt]{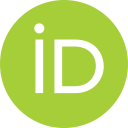}}}

\title[nonthermal effects of Sgr~A* image]{Impact of nonthermal electron radiation effects on the horizon scale image structure of Sagittarius A*}

\author[S. Zhao et al.]{
Shan-Shan Zhao,$^{\orcid{0000-0002-9774-3606},1}$\thanks{E-mail: zhaoss@shao.ac.cn}
Lei Huang,$^{\orcid{0000-0002-1923-227X},1,2}$
Ru-Sen Lu$^{\orcid{0000-0002-7692-
7967},1,3,4}$
and Zhiqiang Shen$^{\orcid{0000-0003-3540-
8746},1,3}$
\\
$^{1}$Shanghai Astronomical Observatory, Chinese Academy of Sciences, Shanghai, 200030, People's Republic of China;\\
$^{2}$Key Laboratory for Research in Galaxies and Cosmology, Chinese Academy of Sciences, Shanghai 200030, People's Republic of China;\\
$^{3}$Key Laboratory of Radio Astronomy, Chinese Academy of Sciences, Nanjing 210008, People's Republic of China;\\
$^{4}$Max-Planck-Institut für Radioastronomie, Auf dem Hügel 69, D-53121 Bonn, Germany.
}

\date{Accepted XXX. Received YYY; in original form ZZZ}

\pubyear{2022}

\begin{document}
\label{firstpage}
\pagerange{\pageref{firstpage}--\pageref{lastpage}}
\maketitle

\begin{abstract}
The Event Horizon Telescope (EHT), with $\sim$20 $\mu$as high angular resolution, recently resolved the millimeter image of the suppermassive black hole in the Galaxy, Sagittarius A*. This opens a new window to study the plasma on horizon scales. The accreting disk probably contains a small fraction of nonthermal electrons and their emissions should contribute to the observed image. We study if such contributions are sufficient to cause structural differences detectable by current and future observational capabilities. 
We introduce nonthermal electrons in a semi-analytical accretion disk, which considers viscosity-leading heating processes, and adopt a continued hybrid electron energy distribution of thermal distribution and power-law tail. We generate the black hole images and extract the structural features as crescent parameters. We find the existence of nonthermal electron radiation makes the crescent much brighter, slightly larger, moderately thicker, and much more symmetric. When the nonthermal connecting Lorentz factor $\gamma_c=65$, which is equivalent to the nonthermal electrons accounting for $\sim1.5$\% of the totals, nonthermal effects cause $\sim2$\% size difference at 230 GHz. Comparing with the structural changes caused by other physical factors, including inclination between the system and the observer, black hole spin, and interstellar medium scattering effects, we find that although nonthermal electron radiation takes the most unimportant role at 230 GHz, it becomes more significant at 345 GHz. 
\end{abstract}




\section{Introduction}

Recently, high resolution millimeter very long baseline interferometry (mm-VLBI) observations \citep[e.g. the Event  Horizon Telescope, EHT, see ][]{EHTC2019ApJL875.L1,EHTC2019ApJL875.L2,EHTC2019ApJL875.L3,EHTC2019ApJL875.L4,EHTC2019ApJL875.L5,EHTC2019ApJL875.L6,EHTC2021ApJ910.L12,EHTC2021ApJ910.L13,EHTC2022ApJ930L12,EHTC2022ApJ930L13,EHTC2022ApJ930L14,EHTC2022ApJ930L15,EHTC2022ApJ930L16,EHTC2022ApJ930L17} and near infrared astrometry \citep[e.g. GRAVITY, see][]{GRAVITY2018AA618.10,GRAVITY2020AA635A.143,GRAVITY2020AA638.A2} open a new window to study the horizon-scaled astrophysics of the Sagittarius A* (Sgr~A*, the suppermassive black hole candidate in the Galactic center). The structure of the black hole image at (sub) millimeter wavelengths becomes detectable and can be used to quantitatively study the physical properties of Sgr~A*, including the spacetime, the magnetic field, the dynamics of the surrounding plasma, and the radiation. 

For radiation, the detected radio waves are emitted from the horizon scale emission region, through the synchrontron radiation mechanism \citep{Yuan2014ARAA52.529}. Then they bend by the curved spacetime in the vicinity of the black hole and finally make up a ring-like bright structure around a dark shadow \citep{Falcke2000ApJ528L13}. Therefore, the magnetic field structure, the electron properties, and the spacetime are all important in modeling the related radiation. Here we focus on the electrons. Suppose all the electrons in the accretion flows are under complete collision, the electrons are in the thermal state. However if some incomplete collision processes exist (e.g. reconnections, shocks), some electrons might be accelerated into higher energy states and become nonthermal. 

Nonthermal electrons are used to explain some observations, for example, adding 1.5\% nonthermal electrons into the thermal electrons can explain the spectrum at low frequency radio band, as well as the x-ray flares \citep{Yuan2003ApJ598.301}. It is also suggested that nonthermal electrons will enlarge the size of Sgr~A*, especially at low frequency \citep{Ozel2000ApJ541.234}. Recently, nonthermal electrons are used to explain new observations such as the Sgr~A* near infrared centroid movement \citep{Petersen2020MNRAS494.5923}, multi-wavelengths flares \citep{Ball2016ApJ826.77,Chatterjee2021MNRAS507.5281,GRAVITY2021AA654A.22} and polarimetry \citep{Mao2018ApJ854.51}. As for black hole images, the ring brightness, size, width and asymmetry are expected to better discriminate the nonthermal electron radiation from the other effects.

\citet{EHTC2022ApJ930L16} considered nonthermal electron radiation as one of the physical inputs to interpret the EHT 2017 data at 230 GHz and found that the result is insensitive to whether use nonthermal electron distribution function (eDF) in their radiation model \citep[consistent with some theoretical expectations that the nonthermal effects are negligible, e.g.][]{Chael2017MNRAS470.2367,Mao2017MNRAS466.4307}. In their works, the nonthermal electrons are mainly introduced in and near the outflow regions, and the results also depend on the electron temperature, which is controlled by a thermodynamical parameter in the relation between the electron temperature and the proton temperature \citep[e.g.][]{Moscibrodzka2016AA586.38,Mizuno2021MNRAS506.741}. This parameter indirectly reveals the physics of the two-tempered accreting plasma (It is generally believed that the electron temperature is much lower than the proton temperature to explain the extremely low bolometric luminosity of Sgr~A*, $L_{\rm bol}\sim 2\times 10^{-9} L_{\rm Edd}$, see, e.g. \citealt{Yuan2014ARAA52.529}).  

In this work, we reconsider the impact of the nonthermal electron radiation on the millimeter black hole image structures. Firstly, we introduce the nonthermal electrons globally into the accretion disk, not the jet/outflow region. 
This is a reasonable choice, since there is no observational conclusion of whether the emissions from the jet region is significant, and the first Sgr~A* horizon scale image resolved by EHT also do not see jet \citep{EHTC2022ApJ930L12}. Secondly, we describe the plasma dynamics using a semi-analytical model by \citet[][here after call the model \textbf{H09}]{Huang2009ApJ703.557,Huang2009ApJ706.960}, where the electron temperature is defined by \cite{Liu2007ApJ668.127}, considering turbulence heating and radiation cooling. There is therefor no need to introduce a free parameter to "post-print" the electron temperature, which simplifies the radiation model. 

Additionally, we extend the analysis from 230 GHz to 345 GHz. The next generation EHT will be improved to 345 GHz \citep[see e.g.][]{Doeleman2019BAAS51.256,Johnson2019BAAS51.235,Raymond2021ApJS253.5}. The higher frequency will not only provide higher resolution, but also reveal the frequency dependent properties in the black hole image. The combination of the current 230 GHz image and the future 345 GHz image is promising in breaking the degeneracy of the effects caused by the nonthermal electron radiation and other physical factors, such as the black hole spin and the interstellar medium scattering.

This paper is written in the following structures. Sect.~\ref{sec:Radiation} shows the nonthermal electron radiation model, which contains a continued hybrid eDF of a thermal distribution with a power-law tail. Sect.~\ref{sec:synthesis_image} shows the method to generate black hole images, containing the dynamical model, the radiation model and the scattering model. We also introduce the method of extracting the crescent structural features from the image domain in this section. In Sect.~\ref{sec:results}, we show the results of nonthermal images, compare the crescent parameters with those extracted from the images of different inclinations, spin and with/without scattering effects, and rank the importance of each physical factors at 230 GHz and 345 GHz. Finally, conclusions and discussions are given in Sect.~\ref{sec:conclusion}.


\section{Radiation model }
\label{sec:Radiation}

The detected radio waves come from synchrotron radiation, which is emitted by the electrons spiralling moving along the magnetic field. The propagation of such radio waves are described by a radiative transfer equation at an observing frequency $\nu$ \citep{Rybicki1979rpabook}:
\begin{equation}
    \frac{d}{d\lambda}\left(\frac{I_\nu}{\nu^3}\right)=\left(\frac{j_\nu}{\nu^2}\right)-(\nu\alpha_\nu)\left(\frac{I_\nu}{\nu^3}\right).
    \label{eq:radiative_transfer}    
\end{equation}
The emissivity $j_\nu$ and the absorption $\alpha_\nu$ are determined by the local electron number density, the magnetic field, the photon moving direction, and also the eDF $n(\gamma)$, which define how many electrons contribute to the radiation at a certain frequency. If assuming all the particles are following complete collision, the electrons are purely thermal and only determined by the temperature. In this situation the eDF is described by the Maxwell distribution. However, if some incomplete collision processes exist, for instance magnetic reconnections and shocks, some electrons are accelerated into higher energy states, which leads the distribution at high energy regime to become a power-law-like shape. 

There are many different ways to build an eDF. The most strict and reasonable way is using particle-in-cell (PIC) simulation with particle acceleration mechanisms \citep[e.g.][]{Lynn2014ApJ791.71,Werner2018MNRAS473.4840,Comisso2019ApJ886.122,Sironi2021ApJ907.44} to compute the local eDF at each grid of emission region. However it is too computationally expansive to realize, so instead, using a simple function to be nonthermal eDF is a more feasible and generally adopted way. For example, the nonthermal eDF can be expressed as a hybrid function of thermal distribution and power-law tail \citep[e.g. ][]{Ozel2000ApJ541.234,Yuan2003ApJ598.301,Ball2016ApJ826.77,Chael2017MNRAS470.2367,Chatterjee2021MNRAS507.5281,EHTC2021ApJ910.L13,EHTC2022ApJ930L16}, kappa distribution\citep[e.g.][]{Pandya2016ApJ822.34,Pandya2018ApJ868.13,Davelaar2019AA632.2,EHTC2022ApJ930L16}, and multi-thermal distributions \citep{Mao2017MNRAS466.4307}.

\begin{figure}
    \centering
    \includegraphics[scale=0.65]{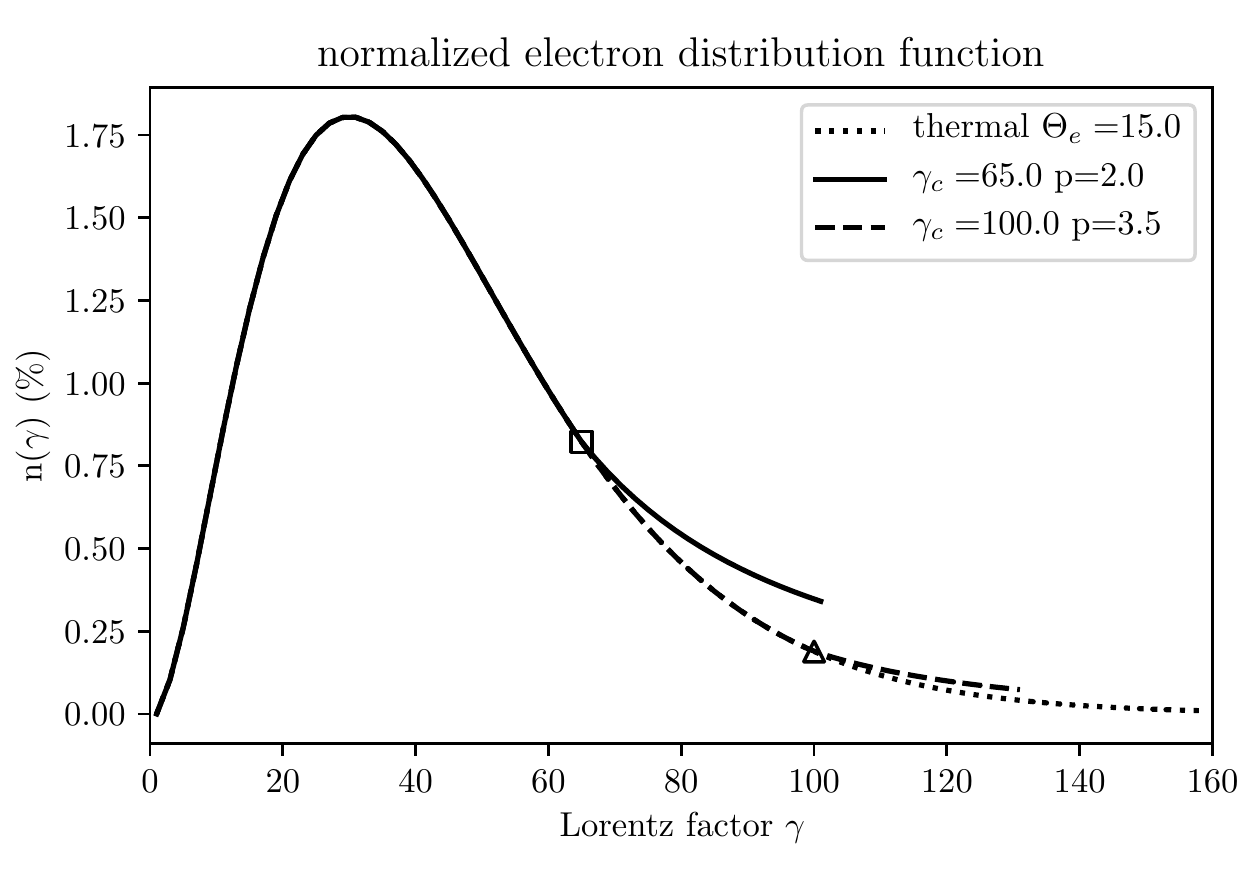}
    \caption{Examples of hybrid normalized eDFs containing a part of Maxwell distribution (thermal electrons) at the low energy region and a power-law tail (nonthermal electrons) at the high energy region. The dimensionless electron temperature is $\Theta_e=15$. We show three cases: the pure thermal eDF (dotted line), the nonthermal eDF with $\gamma_c=65$ and $p=2.0$ (solid line), as well as the nonthermal eDF with $\gamma_c=100$ and $p=3.5$ (dashed line), where the connecting point $\gamma_c=65$ is marked by a square and $\gamma_c=100$ is marked by a triangle. }
    \label{fig:eDF}
\end{figure}

In this work, the adopted eDF is a continuous hybrid function of a part of thermal distribution (the Lorentz Factor $\gamma\le\gamma_c$) and a power-law tail (between $\gamma_c$ and $\gamma_{\rm max}$):
\begin{equation}
n(\gamma)d\gamma=\left\{
            \begin{array}{lr}
            \dfrac{\gamma^2\beta\exp(-\gamma/\Theta_e)}{\Theta_eK_2(\Theta_e^{-1})} d\gamma, & 1\le\gamma\le\gamma_c,  \\[2em]
            C(\gamma_c)\cdot{}\gamma^{-p} d\gamma,& \gamma_c<\gamma\le\gamma_{\rm max},
             \end{array}
\right.
\label{eq:eDF}
\end{equation}
where $\beta=v/c$ is the relative speed of the electron, $\gamma=1/\sqrt{1-\beta^2}$, $\Theta_e=kT_e/m_ec^2$ is the dimensionless electron temperature, $K_\nu(z)$ is the modified Bessel function of the second kind, the connecting point between thermal and power-law distribution is $\gamma_c$ and the power-law index is $p$.  Because of the continuity of $n(\gamma)$ at $\gamma_c$ and the normalization $\int_1^\infty n(\gamma) d\gamma=1$, we have
\begin{equation}
    C(\gamma_c)=\frac{\gamma_c^{p+2}\beta_c\exp(-\gamma_c/\Theta_e)}{\Theta_eK_2(\Theta_e^{-1})},
    \label{eq:eDF_C}
\end{equation}
and
\begin{equation}
        \gamma_{\rm max}\!=\!
         \left\{\left[C^{\!-\!1}\!(\gamma_c\!)\!-\!\frac{\int^{\gamma_c}_1\!\!\gamma^2\beta\exp(-\gamma/\Theta_e)d\gamma}{\gamma_c^{p+2}\beta_c\exp(-\gamma_c/\Theta_e)}\!\right]\!(1\!-\!p)\!+\!\gamma_c^{1\!-\!p}\!\right\}^{\frac{1}{1-p}}.
         \label{eq:eDF_gammac}
\end{equation}
Fig.~\ref{fig:eDF} shows the hybrid eDF with different nonthermal parameters $\gamma_c$ and $p$.

To additionally simplify the eDF, we notice that, $\gamma_c$ is the major indicator of the importance of the nonthermal electrons. If $\gamma_c/\Theta_e\gtrsim 10$, nonthermal electrons are not important in the distribution, we can use pure thermal distribution instead. Observations support that Sgr~A* is dominated by thermal component at millimeter wavelengths, so we assume that at each emission region on the accretion disk, the eDF is always dominated by the thermal distribution. This assumption is realized by constraining the contacting point $\gamma_c$ larger than the peak of the thermal distribution, i.e. $\gamma_c\gtrsim\gamma_{c,\rm min}$, where the lower limit $\gamma_{c,\rm min}$ numerically equals to
\begin{equation}
    \gamma_{c,\rm min}=2\Theta_e+0.25\Theta_e^{-1}.
    \label{eq:gammac_min}
\end{equation}
In order to let $\gamma_{\max}$ have real solution, the power-law index $p$ should not be too large, the numerical solution of the upper limit of $p$ is 
\begin{equation}
    p_{\rm max}\!=\!-0.1057+0.8302(\gamma_c\!/\Theta_e)^{0.1807}
    +1.049\!\exp{\!\left(-1.009\Theta_e\right)}.
    \label{eq:p_max}
\end{equation}
In \textbf{H90}, the electron temperature $\Theta_e$ is relatively small ($\sim$1-20), and $p_{\rm max}$ is also very small. We fix $p$ equals to $p_{\rm max}$ and only let $\gamma_c$ free. For the case of $\gamma_c\sim65$ and $\Theta\sim1$, $p$ is $\sim2$. See Appx.~\ref{sec:app_radiation} for more details. 

By adopting the eDF of Eq.~\eqref{eq:eDF}, the emissivety and the absorption in Eq.~\eqref{eq:radiative_transfer} become $j_\nu=j_\nu^{(1)}+j_\nu^{(2)}$  and $\alpha_\nu=\alpha_\nu^{(1)}+\alpha_\nu^{(2)}$, where the upper script $^{(1)}$ and $^{(2)}$ refer to the thermal part and the nonthermal part. See Eq.~\eqref{eq:j_tp} to \eqref{eq:j_tp_2} and Eq.~\eqref{eq:alpha} to \eqref{eq:alpha_2} in the Appx.~\ref{sec:radiative_transfer}.

We note that $\gamma_c$ is the only free parameter in our nonthermal electron radiation model. When $\gamma_c$ is small, the proportion of nonthermal electrons is large. $\gamma_c$ is independent to $\Theta_e$, which is determined by the accretion flow model.

\section{black hole image generation and feature extraction}
\label{sec:synthesis_image}

The radiation model introduced in Sect.~\ref{sec:Radiation} is numerically realized by ray-tracing method (see Sect.~\ref{sec:ray-tracing}), which traces the light rays bending in the curved spacetime around the black hole, and the corresponding magnetic field and plasma properties at each emitting point come from the dynamical model (see Sect.~\ref{sec:dynamical}). We also use a scattering model (see Sect.~\ref{sec:scattering}) to describe the propagation of the radio waves through the interstellar medium in the Galaxy. We generate millimeter black hole images and extract their structural features (see Sect.~\ref{sec:extraction}).

\subsection{dynamical model: Semi-analytical model H09}
\label{sec:dynamical}
Multi-wavelengths observations of Sgr~A* have shown that it is a hot but quite dim source. This phenomenon can be explained by a geometrically thick but optically thin accretion disk, such as advection-dominated accretion flows \citep[ADAFs, e.g.,][]{Narayan1994ApJ428.13,Moscibrodzka2009ApJ706.497}. The main idea of these models is to introduce a mechanism which is efficient in heating the accretion flow but inefficient in radiation. Thus, this kind of models is also called radiatively inefficient accretion flow (RIAF). We refer to \cite{Yuan2014ARAA52.529} for more details.

There are two kinds of approaches to model the plasma dynamics: one is the semi-analytical approache \citep[e.g.,][]{Huang2009ApJ703.557,Huang2009ApJ706.960,Broderick2011ApJ735.110,Broderick2016ApJ820.137,Pu2016ApJ831.4,Pu2018ApJ863.148}; the other one is general-relativistic magnetohydrodynamical (GRMHD) simulation \citep[e.g.,][]{Noble2007CQG24S.259,Moscibrodzka2009ApJ706.497,Moscibrodzka2016AA586.38,Porth2019ApJS243.26,Mizuno2021MNRAS506.741}. We choose a semi-analytical approach in order to get a concrete electron temperature by solving the electron temperature and proton temperature simultaneously from the MHD equation, which is out of computational feasibility for most of the GRMHD simulations.

\textbf{H09}, the semi-analytical model applied in this work, well describes the electron temperature by assuming the electrons are heated mainly by the viscosity, which is generated by the plasma fluctuations \citep[based on the work by][]{Liu2007ApJ668.127}. Two other processes, the energy exchange between the the electrons and protons, and the radiative cooling effects are also considered. In the model, the density distribution and the velocity filed are obtained by solving the basic equations of the general relativistic accretion flow. The magnetic filed is initialized in a vertical structure. After being sheared by the accretion flow, it is finally in a configuration which is parallel to the velocity field. \textbf{H09} accretion disk is defined by four parameters: the spin of black hole $a_*$, the ratio of the magnetic field energy density to the gas pressure $\beta_p$, dimensionless
constant for electron heating rate $C_1$, constant mass accretion
rate $\dot{M}$. It should be noted that the mass accretion
rate decreases with decreasing radius, due to the existence of strong winds in the accretion flow \citep{Yuan2012ApJ761.130,Yuan2015ApJ804.101}. But considering the wind is weak at the inner region \citep[$r<30r_g$, see][]{Yang2021ApJ914.131}, where the radiation of interest mainly comes from, we treat the mass accretion rate as a constant. We take the models with $a_*=0,\ 0.5,\ 0.9,\ 0.95$ in this work, while other three parameters are chosen carefully in order to make sure the flux density of Sgr~A* at 230~GHz is in the range of $\sim1$ to several Jy with any alternative inclination. Exploring the whole dynamical model parameter space is beyond the scope of this work.

\subsection{ray-tracing scheme: {\tt ipole}}
\label{sec:ray-tracing}

We use the ray-tracing scheme {\tt ipole}\footnote{\href{https://github.com/moscibrodzka/ipole}{https://github.com/moscibrodzka/ipole}} developed by \cite{Moscibrodzka2018MNRAS475.43} to trace the light emitted from the \textbf{H09} accretion disk, bending in the Kerr spacetime and captured by a camera at 1000 $r_g$ far from the black hole. We integrate the radiative transfer equation Eq.~\eqref{eq:radiative_transfer} along each light and finally get the black hole image. It is capable of generating images of any inclination. {\tt ipole} is one of the adopted ray-tracing codes in the theoretical works by EHT collaboration \citep{EHTC2019ApJL875.L5,EHTC2021ApJ910.L13,EHTC2022ApJ930L12}, and have been well compared with other ray-tracing codes by \citet{Gold2020ApJ897.148}. 

\subsection{Scattering: isotropic Gaussian thin screen}
\label{sec:scattering}

At centimeter wavelengths, scattering effects appear as a Gaussian broadening of the source, proportional to the square of the wavelength. At (sub)millimeter wavelengths, diffractive scattering blurs the image by an anisotropic Gaussian, and refractive scattering causes refractive noise on the image. In this work, we focus on ensemble-averaged scattering model, which means that the averaging time is longer than the typical decorrelation timescale for refractive noise, i.e. days to weeks \citep[see][]{Johnson2016ApJ833.74}. Therefore the refractive noise is smoothed, and the scattering is described by an anisotropic Gaussian blur. We use Gaussian parameters measured by \citet{Johnson2018ApJ865.104}: the major axis size $\theta_{\rm maj}=1.38$, minor axis size $\theta_{\rm min}=0.703$, position angle $\phi_{\rm PA}=81.9^\circ$ (reference to the source at 1cm wavelengths). The more complicated scattering models involving refractive scattering effects can be found in the works by e.g. \citet{Johnson2018ApJ865.104,Zhu2019ApJ870.6,Issaoun2019ApJ871.30,Issaoun2021ApJ915.99}, which are beyond the scope of this work.

\subsection{Feature extraction: Parameterize the image into crescent parameters}
\label{sec:extraction}

\begin{figure*}
    \centering
    \includegraphics[scale=0.65]{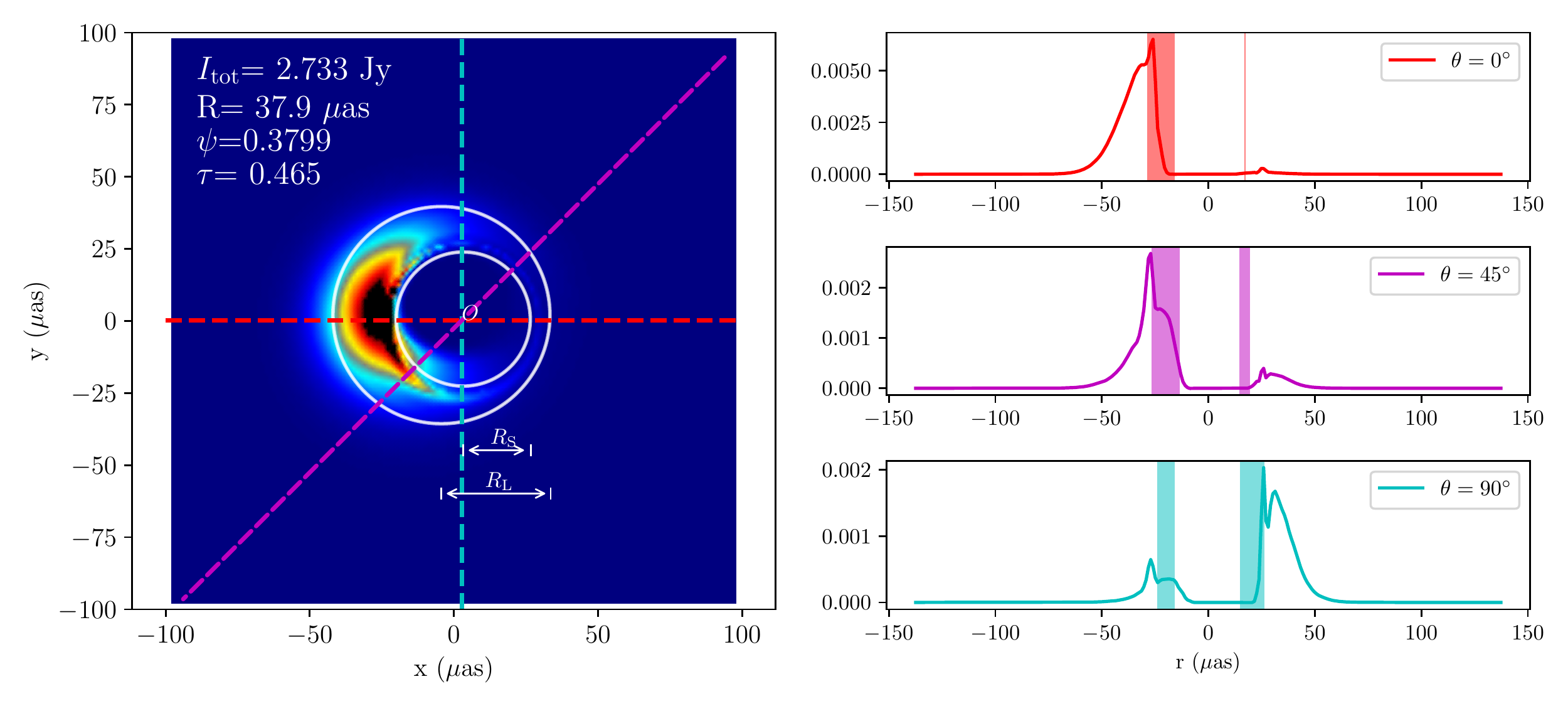}
    \caption{This draft shows how to extract crescent features from a given image. The left panel is the model image of $i=45^\circ$, $a_*=0.5$, thermal model, without scattering effects. A large circle with radius $R_{\rm L}$and a small circle with radius $R_{\rm S}$ are drawn with white color in the image. The red, magenta and cyan dashed lines refer to the angular direction of $\theta=0^\circ$, $45^\circ$, $90^\circ$, and the one dimensional intensity profile at each direction are plotted with the same color in the right panel. The shaded area in the right panel refer to the bright part of the crescent, based on the full width at half maximum of the best-fit Gaussian distributions of the profile. The crescent parameters are in the legend in the upper left corner. }
    \label{fig:method}
\end{figure*}

Inspired by the ring parameter extraction in \citet{EHTC2019ApJL875.L4}, we use crescent model to extract the structural features of a image. A crescent is formed by a large uniformed brightness disk with subtracting a small disk from it, which is described by five parameters: the total intensity $I_{\rm tot}$, the overall size $R$, the relative thickness $\psi$, the degree of symmetry $\tau$ and the orientation $\phi$, following \cite{Kamruddin2013MNRAS434.765}. Extract the crescent parameters from the image domain contain three steps:
\begin{enumerate}
    \item Find the crescent center\\
     The crescent center $(x_0, y_0)$ is defined as the center of the large disk  (where $x_0$, $y_0$ are described in the Cartesian coordinate of the image), and we find it by roughly fitting the image with a infinite-thin-ring. To do the fitting, firstly, we take a polar coordinate with the origin at an arbitrary $(x_0, y_0)$, and then for each position angle $\theta_i$, there is a peak radius $r_{\rm{pk},i}$, which is corresponding to the peak of the one dimensional intensity profile $I(r)|_{\theta_i}$. We change the $(x_0, y_0)$ and finally find the best fit when the peak radius set $\{r_{\rm{pk},i}\}$ has the minimum standard deviation. 
     \label{enum:crescent_parameterize_step1}
\item Find the size of two disks and locate the small disk\\
 We take the crescent center determined by step~\ref{enum:crescent_parameterize_step1} as the origin of the polar coordinate, and for each $\theta_i$, we fit $I(r)|_{\theta_i}$ with Gaussian components to get the locations of their full width at half maximum, where the one closer to the origin is written as $r_{\rm{S},i}$, and the outer one is written as $r_{\rm{L},i}$. We fit the point set $\{(r_{\rm{L},i},\theta_i)\}$ to a circle with radius $R_L$ at the crescent center, while fit $\{(r_{\rm{S},i},\theta_i)\}$ to a smaller circle radius $R_S$, and its center is at $(a,b)$, where $(a,b)$ is described in Cartesian coordinate respect to the crescent center. We take the best fitting $R_L$, $R_S$ and $(a,b)$ as the large disk radius, small disk radius and small disk center of the crescent model, respectively. Fig.~\ref{fig:method} shows an example of feature extraction, where the right panel shows the one dimensional intensity profile at $\theta=0^\circ$, $45^\circ$, $90^\circ$.
\item Calculate the crescent parameters\\
The crescent parameters are defined by \citet{Kamruddin2013MNRAS434.765}:
\begin{eqnarray}
\rm{intensity} && I_{\rm tot}=\sum_{i,j} I_{i,j},\\[0.7em]
\rm{overall\ size} && R=R_L,\\[1.1em]
\rm{relative\ thickness} && \psi=1-\frac{R_S}{R_L},\\
\rm{degree\ of\ symmetry} && \tau=1-\frac{\sqrt{a^2+b^2}}{R_L-R_S}.
\end{eqnarray}
The left panel of Fig.~\ref{fig:method} shows the best fitting crescent and the corresponding parameters. 
\end{enumerate}

\section{Results}
\label{sec:results}

\subsection{Nonthermal images}
\label{sec:results_nonthermal}
 
\begin{figure*}
    \centering
    \includegraphics[scale=0.7,trim=30 0 0 0, clip]{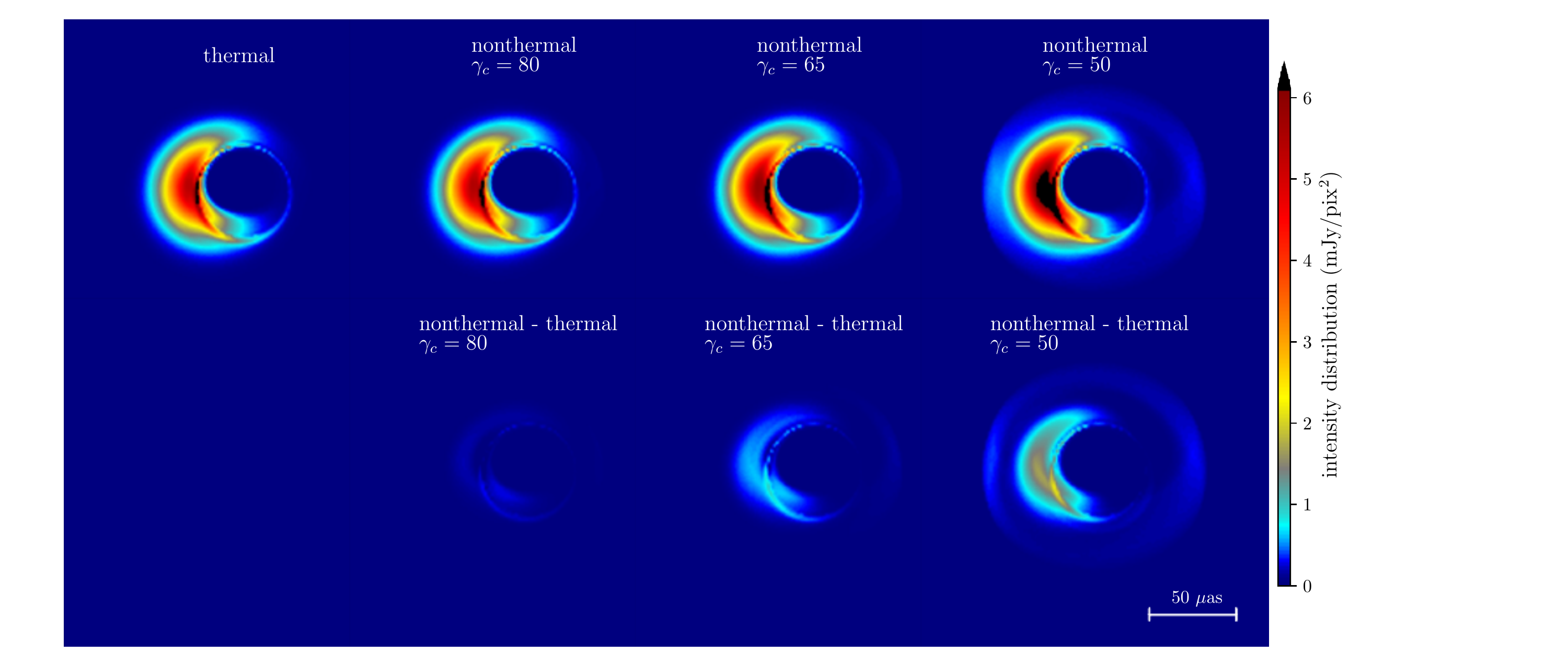}
    \caption{Top row: Black hole images of thermal electron radiation model and the nonthermal electron radiation model with power-law connecting energy $\gamma_c=$80, 65, 50. Bottom row: The remaining intensity distribution of the nonthermal model images after removal of the intensity distribution of the thermal image (the top left). All the images use the same color bar as shown on the right. The observing frequency is 230 GHz, inclination is $i=45^\circ$, spin $a_*=0.5$, without considering scattering effects. }
    \label{fig:nonthermal}
\end{figure*}

Following Sect.~\ref{sec:Radiation} and Sect.~\ref{sec:synthesis_image}, we calculate the black hole images of nonthermal models at 230 GHz (see Fig.~\ref{fig:nonthermal} top row), and their differences from the thermal model (Fig.~\ref{fig:nonthermal} bottom row). For the case of $\gamma_c=80$, it shows almost no difference, while $\gamma_c=50$ shows more differences.

\begin{figure}
    \centering
    \includegraphics[scale=0.55]{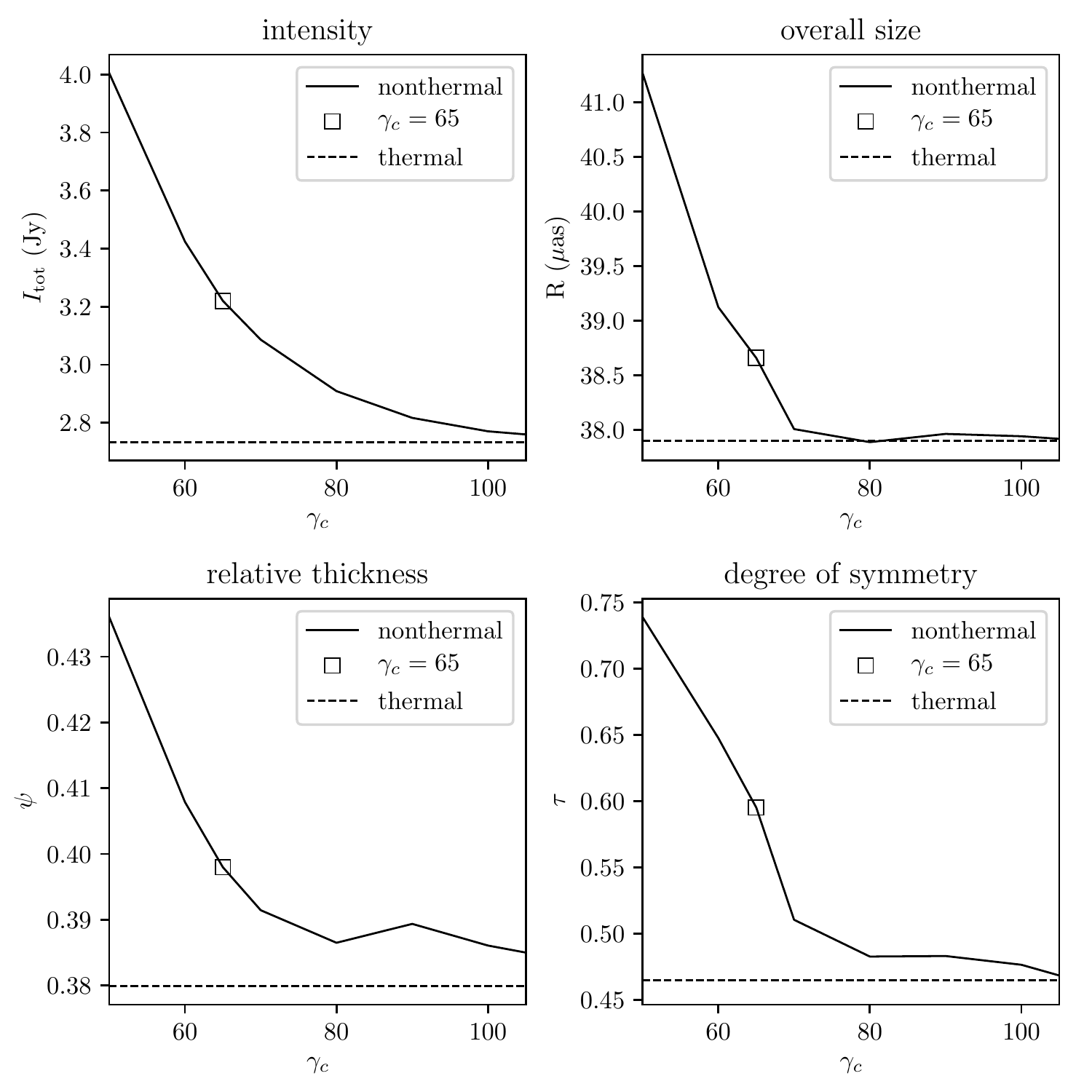}
    \caption{The four crescent parameters extracted from the corresponding black hole images with changing $\gamma_c$ (solid lines), compared with the corresponding values of thermal model (dashed lines). The square maker shows the characterized nonthermal model $\gamma_c=65$, which means that the nonthermal electrons account for $\sim$1.5\% of the totals. The observing frequency is 230 GHz, inclination is $i=45^\circ$, spin $a_*=0.5$, without considering scattering effects.}
    \label{fig:nonthermal_crescent_param}
\end{figure}

To quantify the difference, we extract the structural features of each black hole image with four crescent parameters, i.e. intensity $I_{\rm tot}$, overall size $R_L$, relative thickness $\psi$ and degree of symmetry $\tau$, through the parameterizing method introduced in Sect.~\ref{sec:extraction}. We plot the crescent parameters with varying $\gamma_c$ in Fig.~\ref{fig:nonthermal_crescent_param}. We see that the four crescent parameters of the nonthermal models are all larger than the corresponding parameters of the thermal model, and their differences become small when $\gamma_c$ is large. For a small $\gamma_c$, e.g. $\gamma_c\sim50$, the nonthermal electron radiation increase the intensity from $\sim$2.7 Jy predicted by the thermal model to $\sim$4 Jy, enlarge the overall size from $\sim38\ \mu$as to $\sim42\ \mu$as\footnote{The ring diameter of Sgr~A* measured by EHT is 51.8$\pm$2.3 $\mu$as, and the fractional width is $\sim$30-50 \citep[see Tab.1 in][]{EHTC2022ApJ930L12}, which corresponds to a crescent overall size $R$ in the range of $\sim$39.6-54.1$\mu$as. Considering that our model images are not convoluted with the 20 $\mu$as EHT resolution, we allow $R$ of the model images to be a bit smaller than $39.6\ \mu$as, and only let 54.1 $\mu$as as the upper constraint. }, make it thicker (relative thickness from $\sim$0.38 to $\sim$0.44), and much more symmetric (degree of symmetry from $\sim$0.45 to $\sim$0.75). We also find that the overall size $R$ is only sensitive to small $\gamma_c$, if $\gamma_c\gtrsim 70$, $R$ becomes identical to the value of thermal model. 

The four crescent parameters are all enlarged when $\gamma_c$ is small (small $\gamma_c$ means large nonthermal electron fraction , and their relation is shown in Fig.~\ref{fig:app_radiation}, right panel), because, on the one hand, the outer emission regions that are too dim to be detected in the thermal model case, are now lighted by the radiation emitted from the nonthermal electron. On the other hand, the size and shape of the black hole shadow, which is determined by spacetime, is not affected by the nonthermal model. The removed inner disk remains same, while the outer disk becomes larger and brighter, making the four crescent parameters larger. 

Additionally, we highlight the nonthermal model with a characterized value $\gamma_c=65$ (square markers) in Fig.~\ref{fig:nonthermal_crescent_param}, which is consistent with the $\sim1.5\%$ nonthermal electron fraction suggested by \cite{Yuan2003ApJ598.301}. Sect.~\ref{sec:app_radiation} shows the relation between $\gamma_c$ and the nonthermal electron fraction. We take the model of $\gamma_c=65$ as the representative nonthermal model in the later analysis.

\subsection{Compare with other physical factors: inclination, spin, scattering effects}
\label{sec:results_4factors}

In this section, we quantitatively compare the impact on the black hole image structures caused by the nonthermal electron radiation with other physical factors: the inclination, the spin, as well as the scattering effects. Following Sect.~\ref{sec:results_nonthermal}, we analyze the differences caused by each factor through comparing their crescent parameters. The nonthermal model is chosen as the model of $\gamma_c=65$.

We calculate the 230 GHz image and the 345 GHz image with inclination $i\in(0^\circ,90^\circ)$, spin $a_*=0$, 0.5, 0.9, 0.95, with/without scattering and with/without nonthermal electron radiation. Then we extract the corresponding crescent parameters from those images (see Fig.~\ref{fig:inclination_4x4}).

\begin{table*}
\centering
\begin{tabular}{ccccc|llllllll} 
\hline
\multicolumn{1}{c}{\multirow{2}{*}{model} }      & \multicolumn{1}{c}{\multirow{2}{*}{$i$($^\circ$)} }& \multicolumn{1}{c}{\multirow{2}{*}{$a_*$} } & \multicolumn{1}{c}{\multirow{2}{*}{\begin{tabular}[c]{@{}c@{}}scat.\end{tabular}}}  &
\multicolumn{1}{c|}{\multirow{2}{*}{\begin{tabular}[c]{@{}c@{}}nont.\end{tabular}}}  &
\multicolumn{2}{c}{$I_{\rm tot}$ (Jy)} & \multicolumn{2}{c}{$R$ ($\mu$as)}                                                         & \multicolumn{2}{c}{{$\psi$}}     & \multicolumn{2}{c}{{$\tau$}}      \\ 
\cmidrule(r){6-13}

& &  && \multicolumn{1}{c|}{} & {230 GHz} & {345 GHz} & {230 GHz} & {345 GHz} &{230 GHz} & {345 GHz} & {230 GHz} &{345 GHz}  \\ 
\hline\hline
Mod A&       5&     0.5&        &        &   1.686&   1.029&    42.3&   40.13&  0.4205&  0.3884&  0.9711&   0.969\\
       
Mod B&      45&       0&        &        &    3.24&   2.076&   44.76&    41.8&  0.4889&  0.4254&  0.4679&  0.3496\\

Mod C&      45&     0.5&        &        &   2.733&   2.165&    37.9&   34.77&  0.3799&  0.3089&   0.465&  0.3378\\
      
Mod D&      45&     0.5& $\surd$&        &   2.733&   2.165&   40.13&   36.18&  0.4751&   0.359&  0.3535&  0.2781\\

Mod E&      45&     0.5&        & $\surd$&   3.222&   2.872&   38.66&   36.44&   0.398&  0.3285&  0.5952&  0.5092\\

Mod F&      45&     0.9&        &        &     1.7&   1.445&   32.92&   33.26&  0.2772&  0.2369&  0.1922&  0.1468\\

Mod G&      45&    0.95&        &        &   1.642&   1.625&   31.91&   30.92&  0.2705&  0.2397&   0.221&  0.2199\\

Mod H&      90&     0.5&        &        &   2.844&   3.145&   33.79&    32.4&  0.2558&   0.219&  0.2217&  0.2642\\

\hline
\end{tabular}
\caption{The extracted crescent parameters of the black hole images of different models at 230 GHz and 345 GHz. }
\label{tab:param}
\end{table*}

\begin{figure*}
    \centering
    \includegraphics[scale=0.6]{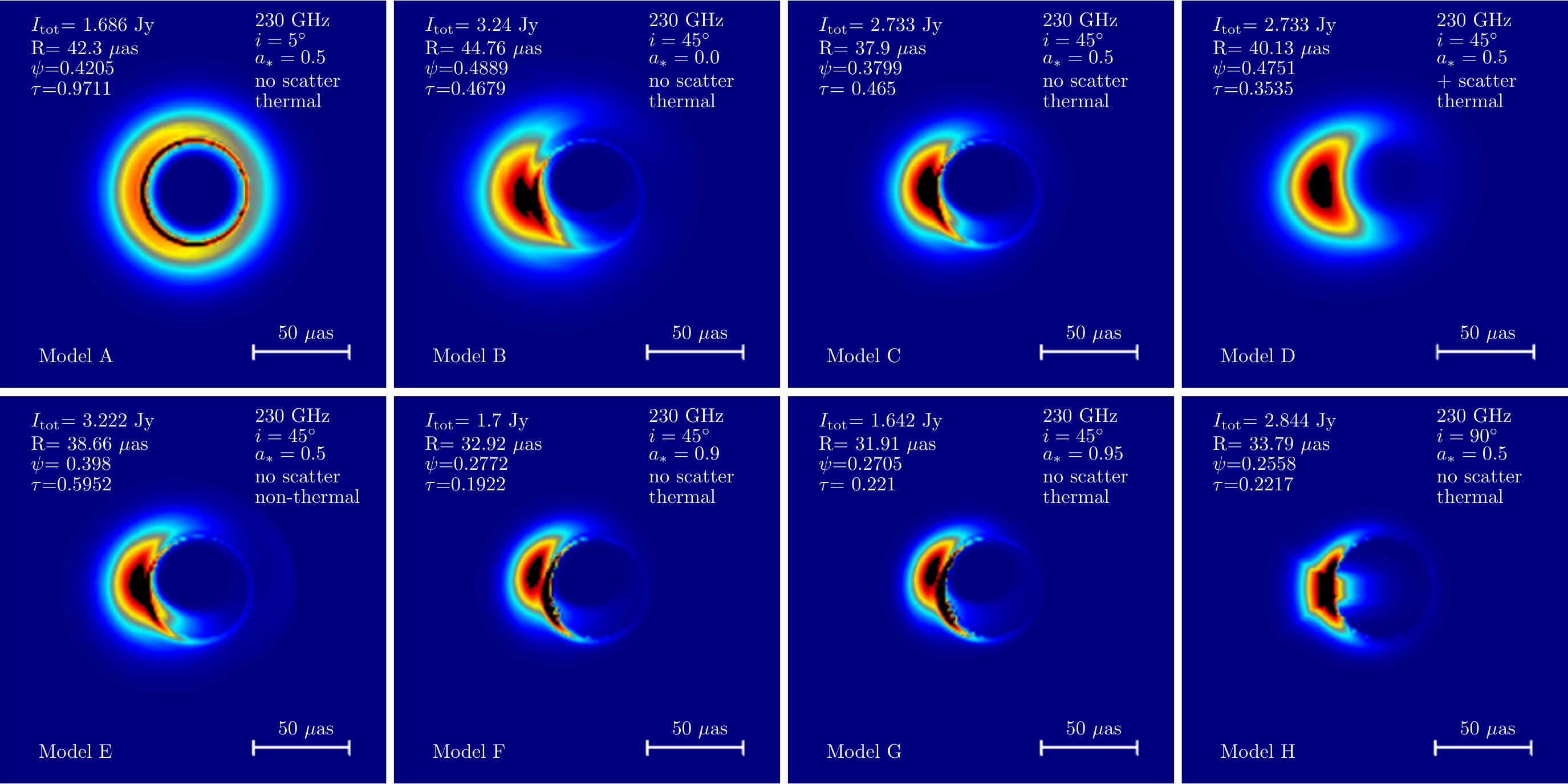}
    \caption{Images of 8 selected models at 230 GHz, with different inclination, spin, with/without scattering and with/without nonthermal electron radiation.}
    \label{fig:display_230}
\end{figure*}

\begin{figure*}
    \centering
    \includegraphics[scale=0.6]{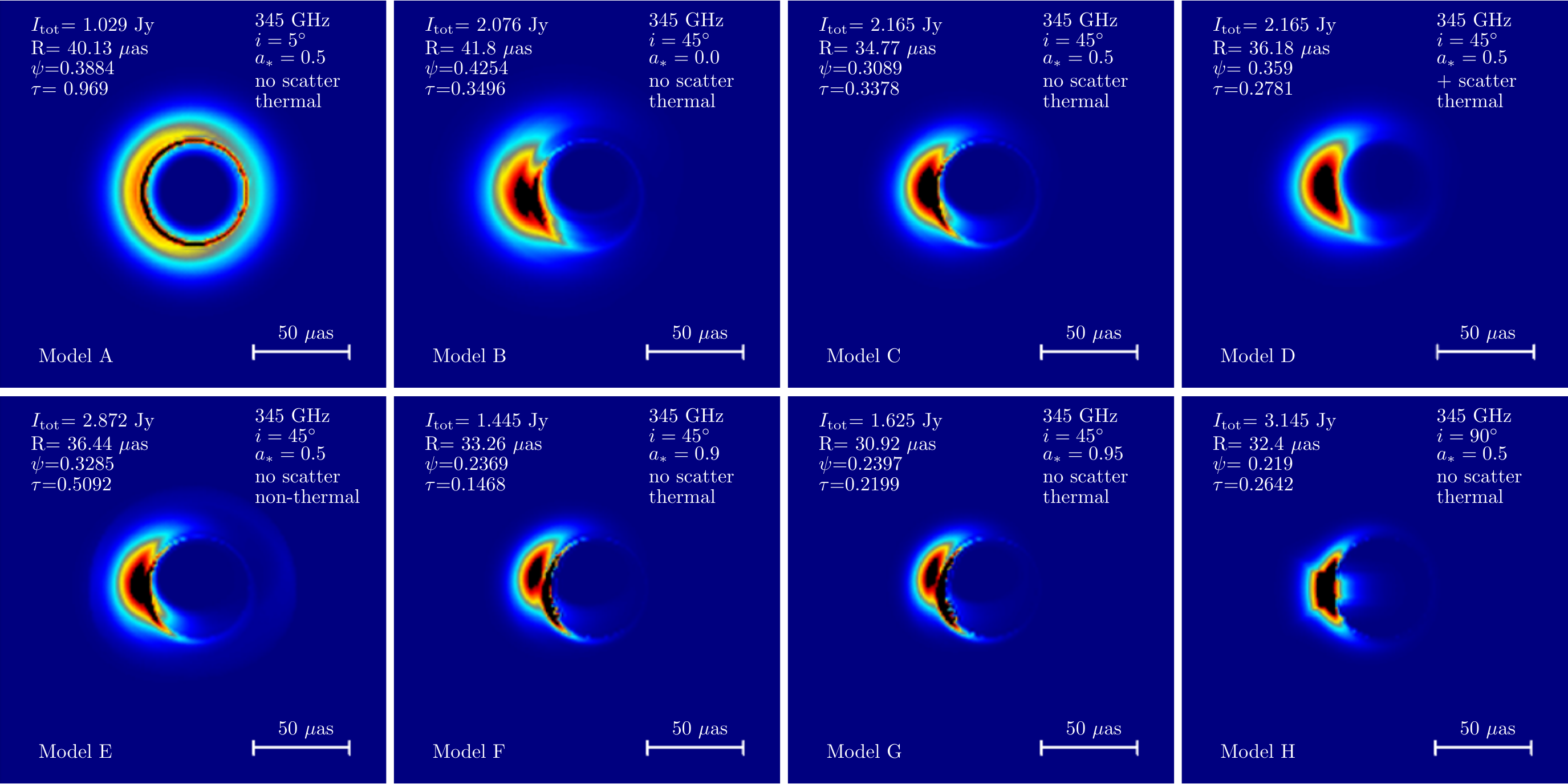}
    \caption{Same as Fig.~\ref{fig:display_230}, but at 345 GHz.}
    \label{fig:display_345}
\end{figure*}

Here we focus on models with $i=5^\circ$, $45^\circ$, $90^\circ$, $a_*=0$, 0.5, 0.9, 0.95, with/without scattering and with/without nonthermal electron radiation. Their crescent parameters are listed in Tab.~\ref{tab:param}. The corresponding images at 230 GHz and 345 GHz are shown in Fig.~\ref{fig:display_230} and Fig.~\ref{fig:display_345}. 

The structural changes of the images due to inclination, spin, scattering effects and the observing frequency are summarised as follows:  
\begin{itemize}
    \item Inclination:
    The face-on-system Model A ($i=5^\circ$) shows ring-like structures with lower intensity, larger size, relatively larger thickness, and high symmetry. Model C($i=45^\circ$) shows highly asymmetric crescent structures and the four crescent parameters changes a lot comparing to Model A, due to the Doppler beaming effects of the rotating plasma. For edge-on Model H ($i=90^\circ$), we find the direct emission from the disk in front of the black hole shadow. We can also see a more compact ring(smaller $R$, $\phi$ and $\tau$), because the strong gravitational lensing effect takes a more important role and concentrates the lights near the photon sphere. 
    
    \item Spin:
    Model F and G refer to fast-rotating black hole ($a_*=0.9$, 0.95). Their images show a much more compact and moderately fainter ring, compared with the image of non-rotating black hole Model B ($a_*=0$). It is because the high spin systems have smaller innermost stable circular orbits (ISCO), which plays the role of the inner cut-off of the accreting plasma. And when the cut-off is closer to the center, the corresponding ultra relativistic plasma near the ISCO have higher energy, which leads to a more concentrate emission distribution. 
    
    \item Scattering effects:
    The model D is same as the model C but considering scattering effects. Scattering effects make the ring larger, thicker and more symmetric, which is similar to the effects caused by the nonthermal electron radiation (see model E). But scattering effects do not change the intensity, while nonthermal model increases the intensity.
    
    \item Frequency:
    Comparing Fig.~\ref{fig:display_230} with Fig.~\ref{fig:display_345}, we find the rings for each model are generally more compact at 345 GHz than 230 GHz. Scattering effects significantly decrease at 345 GHz.  
\end{itemize}


\subsection{Importance of the nonthermal electron radiation at 230/345 GHz}

\begin{figure*}
    \centering
    \includegraphics[scale=0.6]{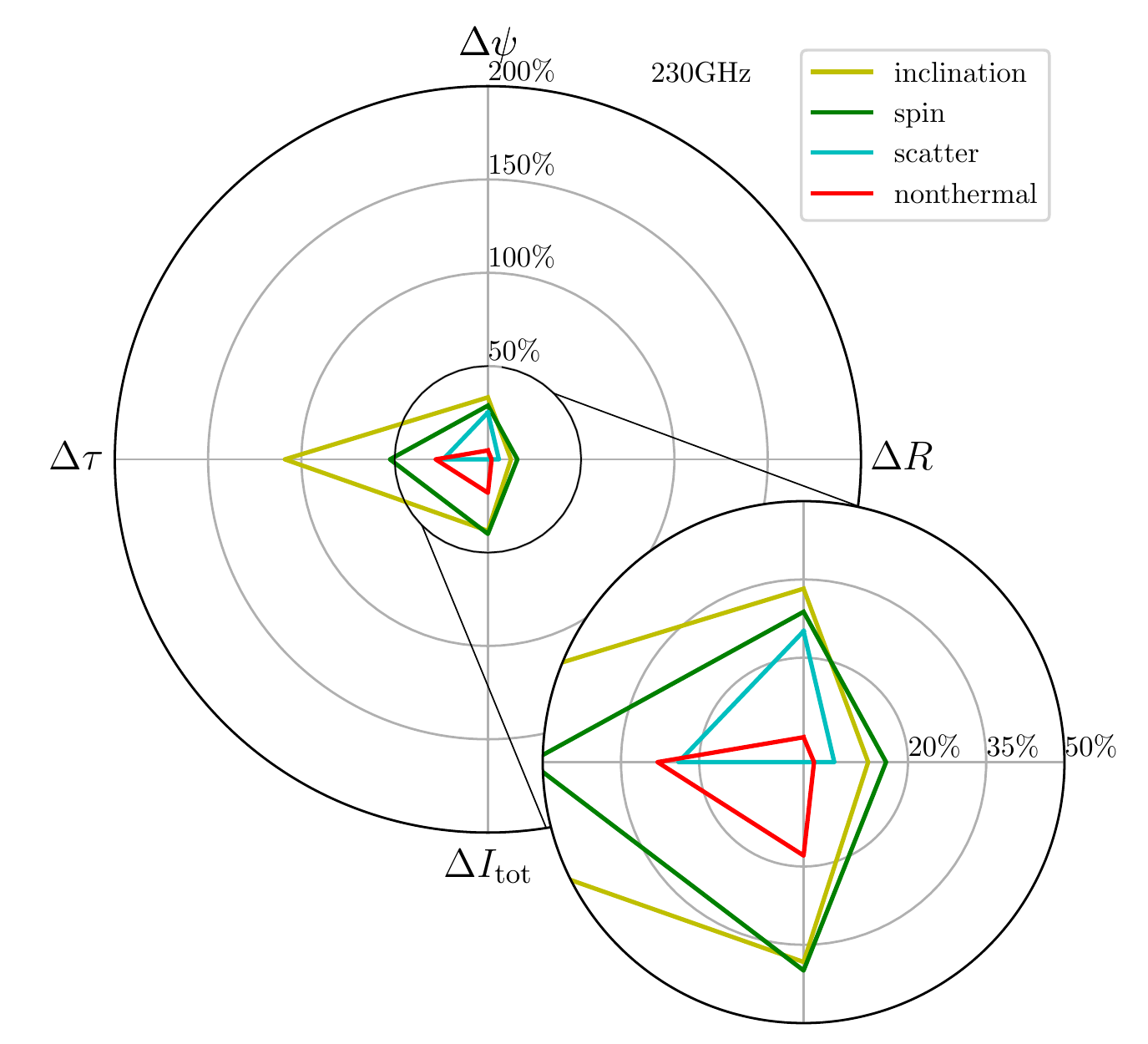}
    \includegraphics[scale=0.6]{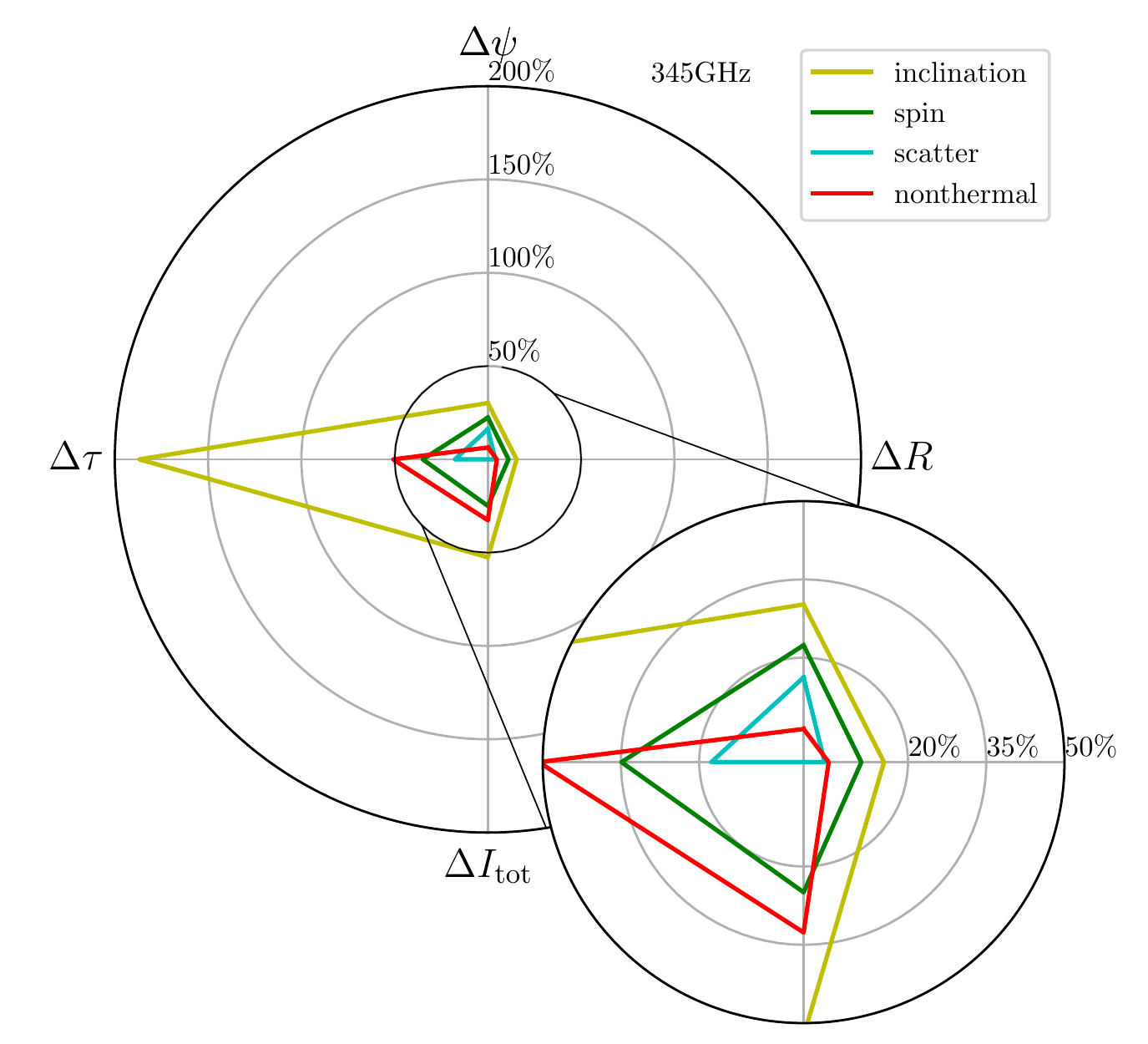}
    \caption{Radar plot of the maximum fractional variations of the crescent parameters due to inclination (olive), spin(green), scattering(cyan), nonthermal electron radiation (red), relative to the fiducial model (model C, $i=45^\circ$, $a_*=0.5$, without scattering effects, without nonthermal electron radiation). The left panel is for 230GHz, the right panel is for 345GHz.}
    \label{fig:radar}
\end{figure*}

To rank the importance of each physical factors, we compare the maximum fractional variations in the crescent parameters, which is relative to the fiducial model (model C, $i=45^\circ$, $a_*=0.5$, without scattering effects, without nonthermal electron radiation). The results are displayed in Tab.~\ref{tab:percentage} and Fig.~\ref{fig:radar}.

\begin{table*}
\centering
\begin{tabular}{c|llllllll} 
\hline
\hline
 & \multicolumn{2}{c}{$\Delta I_{\rm tot}$ (\%)} 
 & \multicolumn{2}{c}{$\Delta R$ (\%)}                                                         
 & \multicolumn{2}{c}{{$\Delta \psi$ (\%)}}     
 & \multicolumn{2}{c}{{$\Delta \tau$ (\%)}}      \\ 
\cmidrule(r){2-9}
&{230 GHz} & {345 GHz} & {230 GHz} & {345 GHz} &{230 GHz} & {345 GHz} & {230 GHz} &{345 GHz} \\ 
\hline\hline
inclination&   38.29&   52.47&   12.37&   15.41&   33.23&    30.2&   108.8&   186.9\\

spin&   39.92&   24.94&   15.81&   11.07&    28.8&   22.39&   52.47&    34.9\\

scatter&       0&       0&   5.882&   4.044&   25.08&   16.23&   23.98&   17.67\\

nonthermal electrons&    17.9&   32.63&   2.001&   4.806&   4.767&   6.351&   27.99&   50.75\\

\hline
\end{tabular}
\caption{The maximum fractional variations of the crescent parameters relative to those extracted from the fiducial model (model C, $i=45^\circ$, $a_*=0.5$, without scattering effects, without nonthermal electron radiation).}
\label{tab:percentage}
\end{table*}

We find that, at 230 GHz, generally speaking, the rank of the importance is: inclination $\gtrsim$ spin $>$ scattering $\sim$ nonthermal. The variations of crescent parameters caused by the nonthermal model are $\Delta I_{\rm tot}\simeq17.9\%$, $\Delta R\simeq2.0\%$, $\Delta \phi\simeq4.8\%$ and the most sensitive one $\Delta \tau \simeq 28.0\%$. The nonthermal electron radiation is the most unimportant factor compared to others, except that the $\Delta I_{\rm tot}$ and $\Delta \tau$ due to the nonthermal model are larger than those due to the scattering effects. However, at 345 GHz, the rank becomes: inclination $>$ nonthermal $\gtrsim$ spin $>$ scattering. The nonthermal variations of the crescent parameters are $\Delta I_{\rm tot}\simeq32.6\%$ (rank 2nd), $\Delta R\simeq4.8\%$ (rank 3rd), $\Delta \phi\simeq6.4\%$ (rank 4th) and $\Delta \tau \simeq 50.8\%$ (rank 2nd). 

Comparing the result at 230 GHz with those at 345 GHz, we summarize: firstly, the structural difference due to the nonthermal model is enhanced at higher frequency. It is because the emission region at 345 GHz is more concentrated in the inner region, where the nonthermal electron radiation is enhanced by gravitational lensing and Doppler boosting. Secondly, the nonthermal electron radiation is the most unimportant physical factor at 230 GHz, while it becomes relatively more important at 345 GHz, due to the enhancement by itself as well as the weakening of scattering effects and spin effects. Since we only consider the Gaussian blur effects in the scattering model, the value is proportional to the square of the observing wavelengths, which makes the scattering effects weaker at 345 GHz. The weakening impact of the spin at 345 GHz is the result of the combined effects of the accretion rate, magnetic field properties and the electron heating process. Finally, between 230 GHz and 345 GHz, the enhancements of the importance of the nonthermal model for the four crescent parameters are not equal: the intensity and the symmetry are significantly increased, while the size and the thickness are only moderately increased.

\section{Conclusions and discussions}
\label{sec:conclusion}
We study the impact of nonthermal electron radiation on the horizon-scaled structures of Sgr~A*. The nonthermal electrons are introduced globally into the accretion disk, which is defined by the semi-analytical dynamical model \textbf{H09}. Their radiation is described by adopting a continued hybrid eDF of a part of thermal distribution with a nonthermal power-law tail, controlled by the connecting point energy $\gamma_c$. We calculate the black hole images at 230 GHz and 345 GHz through ray-tracing scheme, and extract the structural features to four crescent parameters: the intensity $I_{\rm tot}$, the overal size $R$, the relative thickness $\phi$ and the degree of symmetry $\tau$, and then use these crescent parameters to quantify the impact of nonthermal electron radiation. For comparison, we do the same for other physical factors, including the inclination, the spin and the scattering effects. We find that:

\begin{itemize}
    \item the existence of nonthermal electron radiation makes the crescent much brighter, slightly larger, moderately thicker and much more symmetric. Such structural changes are more evident with smaller $\gamma_c$, when $\gamma_c\gtrsim100$, they can be neglected. 
    \item For the case of $\gamma_c=65$ ($\sim$1.5\% of total electrons are nonthermal), the nonthermal model results in a size difference of $\sim$2\% at 230 GHz, which is twice smaller than the uncertainty of ring size measurement by EHT. But at 345 GHz, the size difference increases to $\sim$5\%, which makes it detectable.
    \item The nonthermal electron radiation is the most unimportant factor at 230 GHz, which is comparable to the scattering effects. However it becomes relatively significant at 345 GHz, much more evident than the scattering effects, comparable to the spin and only less important than the inclination.
\end{itemize}

Our 230 GHz result is very consistent with the theoretical interpretation of EHT 2017 observation of Sgr~A* \citep{EHTC2022ApJ930L16} that nonthermal electron radiation is unimportant under the current observational capability. Comparing with the ring fitting results of the EHT data \citep[][]{EHTC2022ApJ930L14}, i.e. $51.8\pm2.3\ \mu$as diameter with $\sim$30\%-50\% fractional width, both of the ring structures of the thermal/nonthermal model images in our work are within this range. The size difference ($\sim$2\%) and thickness difference ($\sim$5\%) due to the nonthermal model are less than the uncertainty of the EHT ring fitting results ($\sim$4\% for size and $\sim$20\% for thickness). 

The future sub-millimeter VLBI observations are very promising in successfully detecting the horizon-scaled structures containing non-negligible nonthermal electron radiation effects. The next generation EHT will provide a factor of two higher angular resolution \citep[from 20 $\mu$as to $10$ $\mu$as, see][]{Doeleman2019BAAS51.256} at 345 GHz, and the $\sim$5\% size difference and the $\sim$6\% thickness difference by the nonthermal model are sufficiently large under such resolution. The asymmetry may also be measured with higher quality data in the future. Our work suggest that the nonthermal model will cause $\sim$50\% asymmetry difference at 345 GHz.

In the more distant future, space VLBI will extend the baseline to a length far more than the earth diameter, and can reach higher frequency since it can avoid atmospheric effects. \cite{Roelofs2019AA625A.124,Palumbo2019ApJ881.62,Johnson2020SciA6.1310,Andrianov2021MNRAS500.4866,Likhachev2022MNRAS511.668} show such ideas and \cite{Johnson2021ApJ922.28} observed Sgr~A* by space VLBI at 1.35 cm. \cite{Chael2021ApJ918.6} suggest that if the angular resolution can be better than $1\mu$as and the dynamic range is sufficiently high, the detection and measurement of inner shadow, or called lensed horizon, is possible. The relative location and the shape of the inner shadow is very sensitive to the inclination, and can be also used to constrain the spin. In addition, the sub-rings, which are the higher order gravitation lensing rings hiding in the unresolved photon ring, may also be used to precisely estimate the spacetime related parameter, such as the spin \citep[see][]{Johnson2020SciA6.1310}. Therefore, because of the lambda-square-scattering effects become very weak at high frequencies, and the inclination and the spin can be constrained through other observables, the nonthermal electron radiation effects may become distinguishable by the space VLBI.  

Our work shows that the the flux density of Sgr~A* increases when the nonthermal model is considered, but this conclusion is based on the time-independent accretion disk. However the flux density of Sgr~A* varies frequently, fluctuating $\sim$1 Jy around 2-3 Jy within several hours at 230 GHz \citep{Wielgus2022ApJ930.19}. The $\sim$18\% intensity difference due to the nonthermal model is smaller than its time variation.

Previous theoretical works predicting the impact of nonthermal electron radiations on millimeter image structures have only focused on 230 GHz. \cite{Chael2017MNRAS470.2367} found that nonthermal image is unchanged when compared to the thermal image, which is consistent with our model's prediction for large $\gamma_c$. \cite{Mao2017MNRAS466.4307} found that the nonthermal model predicts a larger and more diffuse image. This diffusion effect is not seen in our work, because the accretion flow model used in our work does not contain filament structures as theirs. 

Our results depend on the way of introducing the nonthermal electrons. At 230 GHz, the results of introducing nonthermal electrons globally in the accretion disk and introducing the nonthermal electrons mainly in the outflow region\citep{EHTC2022ApJ930L16} are very consistent. It is not clear if the outflow-region-nonthermal model effects also become more evident at 345 GHz or at even higher frequency. If the answer is not, then such two kind of nonthermal models may provide different predictions towards future observations.

\section*{Acknowledgements}

We thank Prof. Yuan, F. of the Shanghai Astronomical Observatory, the internal referee of the Event Horizon Telescope Collaboration and the anonymous referee for helpful comments. This work is financially supported by the China Postdoctoral Science Foundation fellowship (2020M671266); Shanghai Super Postdoctoral Program; Max Planck Partner Group of the MPG and the CAS; the Key Program of the National Natural Science Foundation of China (grant No. 11933007); the Key Research Program of Frontier Sciences, CAS (grant No. ZDBS-LY-SLH011, QYZDJ-SSW-SLH057) and the Shanghai Pilot Program for Basic Research – Chinese Academy of Science, Shanghai Branch (JCYJ-SHFY-2022-013).

\section*{Data Availability}
The data generated in this work will be shared on reasonable request to the corresponding author.



\bibliographystyle{mnras}
\bibliography{example} 




\appendix

\section{Parameter limitations of the radiation model}
\label{sec:app_radiation}

\begin{figure*}
    \centering
    \includegraphics[scale=0.6]{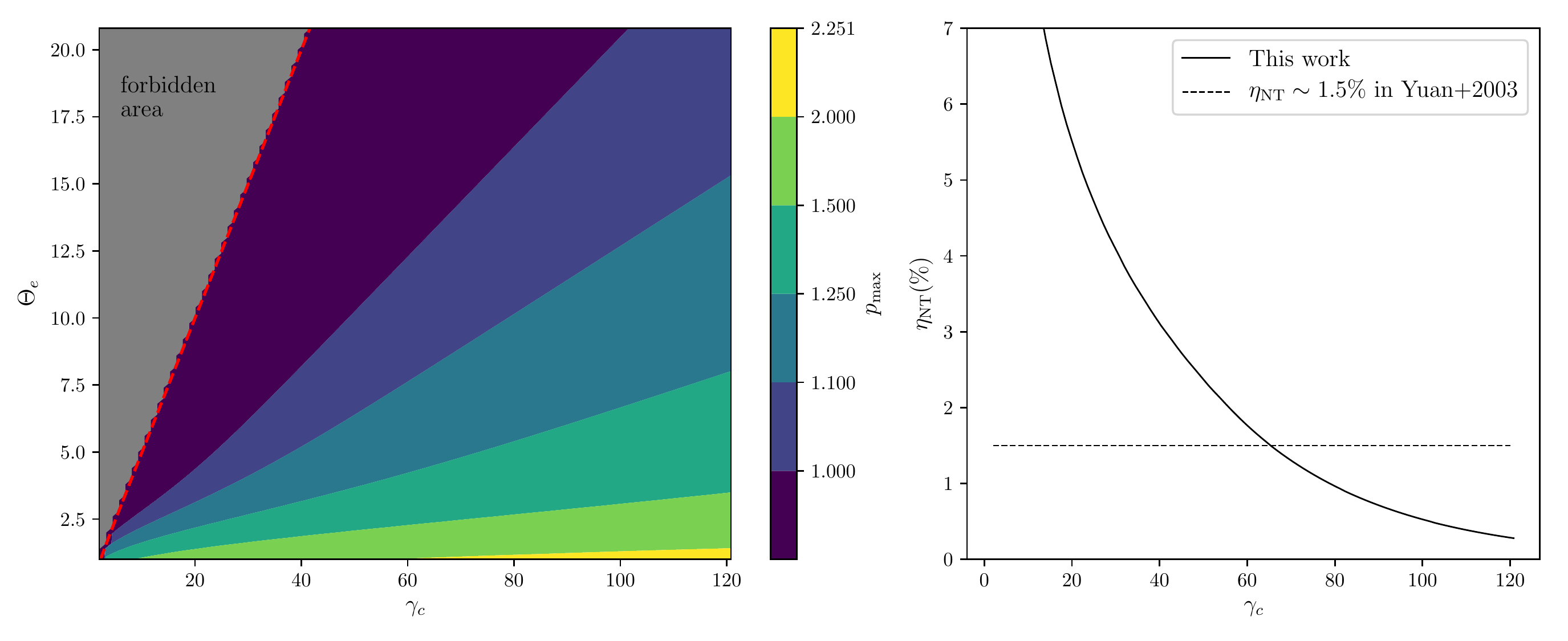}
    \caption{Left: the possible parameter space of the connecting point $\gamma_c$ and power-law index $p$ with different electron temperature $\Theta_e$. The color represents the max value of $p$ as defined by Eq.~\eqref{eq:p_max}. The red dashed line shows the minimum $\gamma_c$ which is defined by Eq.~\eqref{eq:gammac_min}. The forbidden area colored by grey is where $\gamma_c<\gamma_{c,\rm min}$. Right: Averaged fraction of the nonthermal electrons in the accretion disk $\eta_{\rm NT}$ with different $\gamma_c$ (solid line). Comparing with the suggested value $\eta_{\rm NT}\sim 1.5\%$ in the work of \citet{Yuan2003ApJ598.301} (dashed line), we find that $\gamma_c$ is $\sim$65. }
    \label{fig:app_radiation}
\end{figure*}

The nonthermal radiation model shown in Sect.~\ref{sec:Radiation} has two parameters: the connecting Lorentz factor $\gamma_c$ and the index $p$. We consider the following three constraints of the parameter space:
\begin{enumerate}
    \item We compare the emissivity Eq.~\eqref{eq:j_tp} with the approximate emissivity of the thermal model \citep[given by][]{Mahadevan1996ApJ465.327}, and find that when $\gamma_c/\Theta_e\gtrsim 10$, their difference is less than $\mathcal{O}(10^{-2})$, which means that we can use thermal eDF instead of nonthermal eDF. In \textbf{H09}, the majority part of the accretion disk has the electron temerarure $\Theta_e \lesssim 10$, so if $\gamma_c\gtrsim 100$, the nonthermal electron radiation contribute very little to the image.
    
    \item The eDF should be thermal-dominated and the nonthermal part should not be too large.
    We let the connecting point $\gamma_c$ larger than the peak position $\gamma_{\rm pk}$ of the thermal distribution. $\gamma_{\rm pk}$ is defined by $[\partial n_{\rm th}(\gamma)/\partial \gamma]|_{\gamma_{\rm pk}}=0$, which is equivalent to
    \begin{equation}
    \left(2\gamma^{-1}+\gamma^{-3}\beta^{-2}-\Theta_e^{-1}\right)\big{|}_{\gamma_{\rm pk}}=0.
    \end{equation}
    When $\gamma_{\rm pk}$ is larger than 1, the numerical solution is
    \begin{equation}
    \gamma_{\rm pk}=2\Theta_e+0.25\Theta_e^{-1}.
    \end{equation}
    Considering $\gamma_{\rm c,min}=\gamma_{\rm pk}$, the above equation becomes Eq.~\eqref{eq:gammac_min}.
    
    \item In order to let Eq.~\eqref{eq:eDF_gammac} have real solution, there is an inequality depending on the dimensionless electron temperature $\Theta_e$, $p$ and $\gamma_c$. We let $\Theta_e$ and $\gamma_c$ free, and solve the upper limit $p_{\rm max}$ for each $\Theta_e$ and $\gamma_c$, and then plot the results in the left panel of Fig.~\ref{fig:app_radiation}. It seems that $p_{\rm max}$ depends on $\gamma_c/\Theta_e$, and this relation changes when $\Theta_e$ is small. So fitting the data in the plot, we find a numerical approximation of $p_{\rm max}$, which is Eq.~\eqref{eq:p_max}. 
\end{enumerate}
 
 Additionally, when $p=p_{\rm max}$ is fixed, we obtain the relation between our model parameter $\gamma_c$ and the fractional number of the nonthermal electrons $\eta_{\rm NT}$ by integrating the nonthermal eDF. Since different emission regions in the accretion disk have different electron temperatures, we average the results of each region. We plot the $\eta_{\rm NT}$ with different $\gamma_c$ in the right panel of Fig.~\ref{fig:app_radiation}. Then we read that when $\eta_{\rm NT}\sim 1.5\%$, the corresponding $\gamma_c$ is about 65.

\section{Radiative transfer}
\label{sec:radiative_transfer}

The covariant equivalence radiative transfer equation at observed frequency $\nu$ is Eq.~\eqref{eq:radiative_transfer}.
For isotropic electrons, the emissivity $j_\nu$ at emitted frequency $\nu$ comes from the overall contributions of every single electron with energy $\gamma$:
\begin{equation}
    j_\nu=\frac{1}{4\pi}\int^\infty_1 d\gamma N_e n(\gamma) P(\nu,\theta_B,\gamma),
    \label{eq:j_def}
\end{equation}
where $N_e$ is the electron number density.
In the ultra relativistic scenario, i.e. $\gamma\gg1$, the single electron emission power is \citep[e.g.][]{Rybicki1979rpabook,Mahadevan1996ApJ465.327} 
\begin{equation}
    P(\nu,\theta_B,\gamma)=\frac{\sqrt{3}e^3B\sin{\theta_B}}{m_ec^2}F\left(\frac{\nu}{\nu_c}\right),
    \label{eq:j_single_energy}
\end{equation}
where $\theta_B$ is the angle between the wave-vector and the magnetic field $\Vec{B}$, $B$ is the magnetic flux density, $e$ is the electron charge, $m_e$ is the electron mass, $c$ is the light speed,
\begin{equation}
    \nu_c=\frac{3}{2}\nu_B\sin{\theta_B}\gamma^2,
\end{equation}
is the characteristic frequency and
\begin{equation}
    \nu_B=\frac{eB}{2\pi m_e c},
\end{equation}
is the cyclotron frequency. $F(x)$ is a integral function of the second kind modified Bessel function, which is
\begin{equation}
    F(x)=x\int^\infty_x K_{5/3}(y)dy.
\end{equation}
We take $n(\gamma)$ defined in Eq.~\eqref{eq:eDF} into Eq.~\eqref{eq:j_def}, and get
\begin{equation}
    j_\nu=j_\nu^{(1)}+j_\nu^{(2)},
    \label{eq:j_tp}
\end{equation}
\begin{equation}
    j_\nu^{(1)}=\dfrac{N_e e^2 \nu}{\sqrt{3}cK_2(\Theta_e^{-1})x_M}
    {\displaystyle \int_{\Theta_e^{-1}}^{\gamma_c\Theta_e^{-1}} dz\  z^2\exp(-z)F\left(\dfrac{x_M}{z^2}\right)},
    \label{eq:j_tp_1}
\end{equation}
\begin{equation}
    j_\nu^{(2)}=\dfrac{N_e e^2\nu_pC(\gamma_c)}{2\sqrt{3}c}\left(\dfrac{\nu}{\nu_p}\right)^{\frac{1-p}{2}}{\displaystyle \int^{\frac{\nu}{\nu_p}\gamma_{c}^{-2}}_{\frac{\nu}{\nu_p}\gamma_{\rm max}^{-2}}dx\ x^{\frac{p-3}{2}}F(x)},
    \label{eq:j_tp_2}
\end{equation}
where $j_\nu^{(1)}$ and $j_\nu^{(2)}$ represent the emissivity of the thermal and the nonthermal parts. $x_{\rm M}\equiv2\nu/(3\Theta_e^3\nu_B\sin{\theta_B})$ and $\nu_p=3/2\nu_B\sin{\theta_B}$. If $\gamma_c/\Theta_e\gtrsim 10$ and $\Theta_e\gg 1$, we use the approximated thermal emissivity Eq.~A(12) in \cite{Dexter2016MNRAS.462.115}  \citep[firstly given by][]{Mahadevan1996ApJ465.327} instead.

The absorption coefficient is defined as \citep{Rybicki1979rpabook}
\begin{equation}
    \alpha_\nu=\frac{c^2}{8\pi h\nu^3}\int d\gamma \gamma^2 N_e \left[\frac{n(\gamma^*)}{\gamma^*{}^2}-\frac{n(\gamma)}{\gamma^2}\right]P(\nu,\theta_B,\gamma),
    \label{eq:alpha_def}
\end{equation}
where
\begin{equation}
    \gamma^*=\gamma-\frac{h\nu}{m_e c^2}.
\end{equation}
We take the $n(\gamma)$ by Eq.~\eqref{eq:eDF} and the emitted power of the single energy electron by Eq.~\eqref{eq:j_single_energy} into Eq.~\eqref{eq:alpha_def} and have
\begin{equation}
    \alpha_\nu=\alpha_\nu^{(1)}+\alpha_\nu^{(2)},
    \label{eq:alpha}
\end{equation}
\begin{equation}
    \alpha_\nu^{(1)}=\frac{j_{\nu}^{(1)}}{B_\nu(\Theta_e)},
    \label{eq:alpha_1}
\end{equation}
\begin{equation}
    \alpha_\nu^{(2)}=\frac{N_e e^2(p+2)}{4\sqrt{3} m_ec\nu_p}C(\gamma_c)\cdot\left(\dfrac{\nu}{\nu_p}\right)^{-\frac{p+4}{2}}{\displaystyle \int_{\frac{\nu}{\nu_p}\gamma_{\rm max}^{-2}}^{\frac{\nu}{\nu_p}\gamma_c^{-2}}dx\ x^{\frac{p}{2}-1}F(x)}.
    \label{eq:alpha_2}
\end{equation}
$\alpha_\nu^{(1)}$ and $\alpha_\nu^{(2)}$ refer to the absorption coefficient of thermal and nonthermal parts. The thermal part Eq.~\eqref{eq:alpha_1} means the thermal electrons are in the state of complete collision, so their thermodynamics are following the local thermodynamic equilibrium, which means the absorption coefficient and the emissivity can be related by a Boltzmann distribution $B_v(\Theta_e)$. It is also called Kircoff's law.

The double integrals in $j_\nu^{(2)}$ and $\alpha_\nu^{(2)}$ brings difficulties in directly computing the radiative transfer coefficients. Some works adopt approximations to replace the exact solution, \cite[e.g.][]{Huang2011MNRAS416.2574,Dexter2016MNRAS.462.115,Pandya2018ApJ868.13}. Our solution is searching the interpolation from a number list which have saved the results of the double integrals with a grid of different $p$ and the lower and upper limits of each integral. The precision of the interpolated number is $\lesssim 10^{-3}$, which is enough for this work.
\section{Full results}

\begin{figure*}
    \centering
    \includegraphics[scale=0.7]{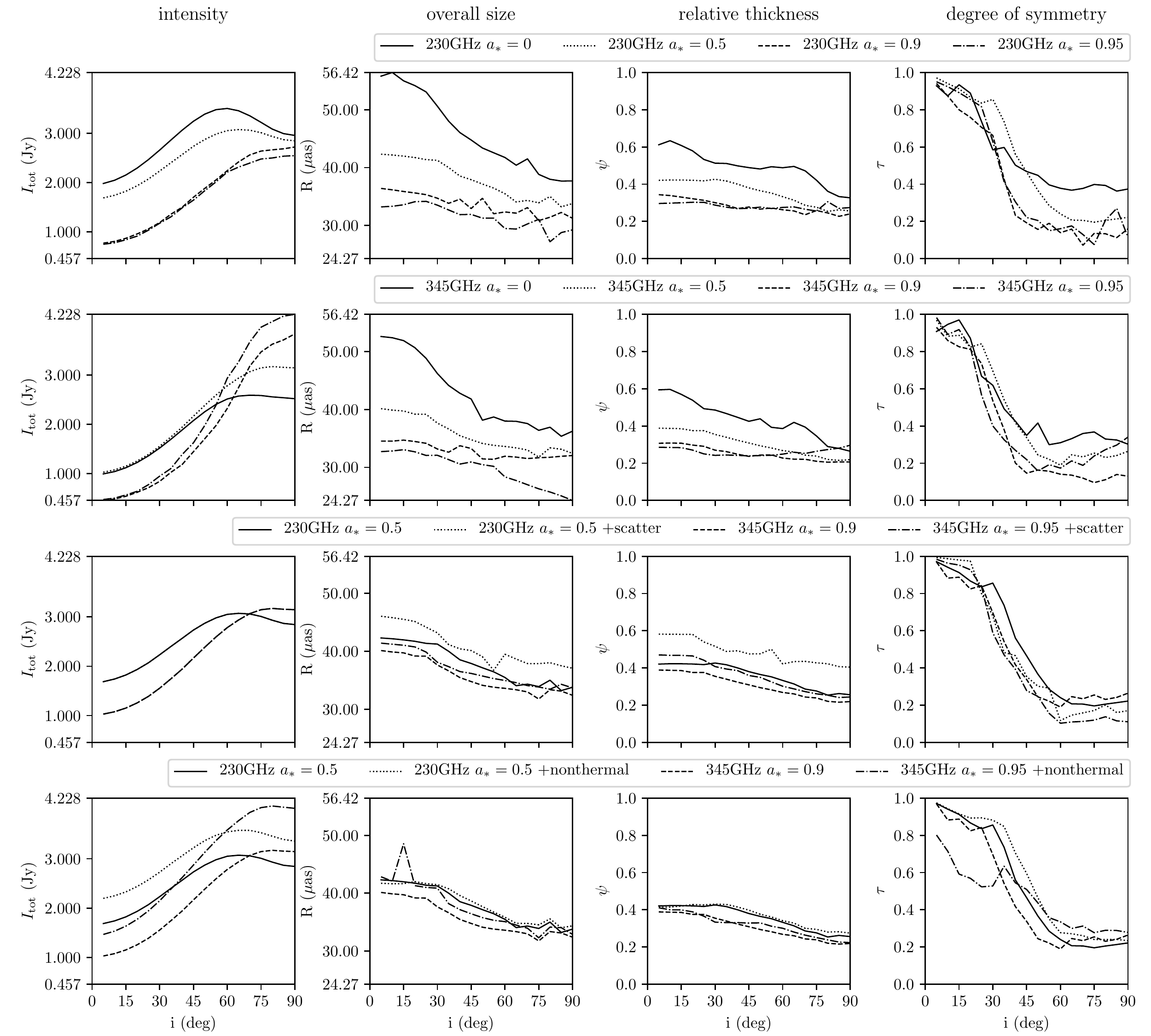}
    \caption{The four crescent parameters extracted from black images of inclination between $0^\circ$ and $90^\circ$. Top row: spin $a_*=$0, 0.5, 0.9, 0.95. (plotted with solid, dotted, dashed and dash-dotted lines) at 230 GHz. Second row: same with the top row but for 345 GHz. Third row: $a_*=$0.5, 230GHz with/without scattering, 345GHz with/without scattering. Bottom row: the same as the third row but for with/without nonthermal electron radiation. }
    \label{fig:inclination_4x4}
\end{figure*}

We extract crescent parameters of the 230 GHz image and the 345 GHz image with inclination $i\in(0^\circ,90^\circ)$, spin $a_*=0$, 0.5, 0.9, 0.95, with/without scattering and with/without nonthermal electron radiation. Sect.~\ref{sec:results_4factors} only shows 8 examples, while the full results are ploted in Fig.~\ref{fig:inclination_4x4}.

\bsp	
\label{lastpage}
\end{document}